\documentclass[11pt]{article}
\usepackage{amssymb,amsmath,amsthm,amsfonts,amscd}
\usepackage{graphicx}
\usepackage{here}
\newcommand\be{\begin{equation}}
\newcommand\ee{\end{equation}}
\newcommand\bea{\begin{eqnarray}}
\newcommand\eea{\end{eqnarray}}

\newcommand{\fatalpha}{{\bf \alpha \kern -0.44em \alpha}}
\newcommand{\fatsigma}{{\bf \sigma \kern -0.54em \sigma}}
\newcommand{\tpchi}{{\bf \chi \kern -0.35em \chi}}
\newcommand{\llambda}{{\bf \lambda \kern -0.45em \lambda}}

\bibliography{plain}
\title{\bf Anti-symmetry consideration on the preservation of Entanglement of spin system}\vspace{20mm}
\author{
  M. A. Fasihi
 \thanks{E-mail: a.fasihi@gmail.com,\quad ma-fasihi@azaruniv.edu}
\\  {\small Department of Physics, Azarbaigan Shahid Madani University, 53714-161 Tabriz, Iran}} \pagebreak

\vspace{20mm}
\begin{document}
\maketitle \vspace{15mm}
\abstract{In this work we offer an approach to protect the entanglement based on the anti-symmetric property of the hamiltonian. Our main objective is to protect the entanglement of a given initial three-qubit state which is governed by hamiltonian of a three-spin Ising chain in site-dependent transverse fields. We show that according to anti-symmetric property of the hamiltonian with respect to some operators mimicking the time reversal operator, the dynamics of the system can be effectively reversed. It equips us to control the dynamics of the system. The control procedure is implemented as a sequence of cyclic evolution; accordingly the entanglement of the system is protected for any given initial state with any desired accuracy an long-time. Using this approach we could control not only the multiparty entanglement but also the pairwise entanglement.
It is also notable that in this paper although we restrict ourselves mostly within a three-spin Ising chain in site-dependent transverse fields, our approach could be applicable to any n-qubit spin system models.

}
\section{Introduction}

Among all of the strange aspects of quantum theory, which never occurs in its classical counterpart,
the most mysterious one is entanglement. Study of the dynamics of entanglement provide the
fascinating insight into our understanding about new emerging quantum technologies.
In fact entanglement is the fundamental resource for quantum computation and quantum
information processing\cite{Nielsen,Gruska,Nakahara}. While the entanglement provides undeniable
advantages to achieve quantum computers its considerable sensitivity to system-environment
interactions, which induces decohering effects in the quantum evolution,
is the main impasse to preserve entanglement. Therefor to struggle against
the effect of decoherence and to control and maintain the entanglement, various strategies have been proposed,
e.g., quantum error correcting codes\cite{Cirac,Calderbank,Knill,Lidar,Lidar book}, decoherence-free subspaces\cite{Lidar1,Mohseni,Mundarain}, weak measurements\cite{Koashi,Korotkov,Kim,Kim1,Kim2}, quantum zeno and super zeno effect \cite{Facchi,Facchi1,Maniscalco,Rossi,Deepak} and quantum bang-bang control\cite{Lloyd,Lloyd1,Vitali}. Inspired by refocusing techniques from NMR \cite{Levitt} Vitali et al. \cite{Vitali} put forward an interesting practical scheme to control and suppress decoherence using tailored external forcing acting as pulses. They showed that provided that the system Hamiltonian possesses
suitable symmetry with respect to parity, undesired effect of the environment such as dissipation, decoherence,
could be effectively averaged out.
Following the idea of parity kick technique, Morigi et al. \cite{Morigi} proposed a scheme which determines
the contributions from coherent and incoherent processes in the evolution
of the system. In summary, to this end, they consider a free evolution of the system for
the time $T$ and then apply a short electromagnetic pulse leading to a
reversal of the system’s unitary evolution. Accordingly at time 2T the
system is expected to returns to its initial state if there is no decoherence.
Following the directions of \cite{Morigi} Meunier et al. \cite{Meunier} investigate
coherent atom-field processes in cavity quantum electrodynamics.
Recently, Rossi \cite{Rossi1} investigate entanglement preservation of two coupled modes as an alternative application of quantum zeno effect.
On the other hand Hou and et al. \cite{Hoa} introduced an interesting scheme based on quantum zeno effect to preserve the entanglement of any given maximally entangled two qubit initial state by controlling the dynamics of the system which is governed by two-qubit Heisenberg model
with DM interaction. The experimental verification of this work is appeared in \cite{Manu}.
In a very recent work the preservation of the long-time limit of entanglement between two qubits via the addition of qubits was
presented \cite{Flores}. In this contribution, considering parity kick method \cite{Vitali} and motivated by the work of Hou and et al. \cite{Hoa} we proposed an approach to preserve the multiparty and pairwise entanglement of three-qubit GHZ state which is governed by three-particle Ising model with transverse field, the shortest nontrivial chain. It is also worth to mention that generally our proposal could effectively be applied to preserve the entanglement of $n$-qubit state which is governed by $n$-particle spin system. The paper is organized as follows. Section 2 introduces our general idea of cyclic evolution based on the antisymmetric properties of hamiltonian and then the measure of concurrence vector is utilized to quantify the multiparty and pairwise entanglement. As an application and efficiency of the approach the dynamics of multiparty and pairwise entanglement of three-qubit GHZ state which is governed by three-particle Ising model with transverse field have been compared with their free evolution dynamics. Section 3 is devoted to a brief conclusion.

\section{General Idea}

Let us consider N-qubit Ising model with transverse field described with the following Hamiltonian
\be\label{Ising1}
H=\sum_{i=1}^{N-1}{J_i \sigma_{i}^z\sigma_{i+1}^z} +\sum_{i=1}^{N}{ h_i \sigma_{i}^x}
\ee
where $J_i$'s are the coupling constants and $\sigma_{i}^x$ and $\sigma_{i}^z$ denote the $x$ and $z$ components of Pauli spin matrices at the $i$th site respectively.
Our purpose is to protect the entanglement of a given initially maximal entangled state, which is governed by hamiltonian of Eq.(\ref{Ising1}), by controlling the dynamics of the system.
To this end, inspired with parity kick method \cite{Vitali}, we put forward our control scheme based on the following necessary condition :

\textit{The hamiltonian $H$ must be antisymmetric with respect to some operators, in other words, it must be at least a hermitian unitary operator, say $A$, which is anti-commuting with $H$.}
\be\label{Antisymmetry}
H A = -A H~~~  \Rightarrow ~~  A e^\frac{-iH t}{\hbar} = e^\frac{iH t}{\hbar}A~~  \Rightarrow ~~ A e^\frac{-iH t}{\hbar}A = e^\frac{iH t}{\hbar},
\ee
 This condition plays the crucial rule in our scheme. The last identity in the above equation shows that the operators $A$ is mimicking the time reversal operator. Now assuming that the antisymmetric condition is held, to complete our scheme we have to consider the following four steps.
\begin{itemize}
  \item Firstly we let the system evolves from time $t=0$ to $t=T$ under
         the action of the unitary evolution operator $ U(t)=e^{\frac{-i H t}{\hbar}}$
  \item Secondly at time $t=T$ we apply the unitary operator $A$
  \item Thirdly for the remaining time $t-T$ we apply the time evolution operator $ U(t-T)=e^{\frac{-i H (t-T)}{\hbar}}$
  \item   Finally as a forth step, at $t=2T$ we again apply operator $A$.
\end{itemize}
Then the overall time evolution operator is a cyclic
evolution with a period of $2T$ :
\be\label{cyclic1}
U_{cycl}(2T)=A e^{\frac{-i H (2T-T)}{\hbar}}A e^{\frac{-i H T}{\hbar}}=I,
\ee

where I stands for Identity operator.
Clearly at the times $t=T$ and $t=2T$ the cyclic evolution is not continuous.
Note that, the last identity of above cyclic
evolution with a period of $2T$, wouldn't be achieved if the the antisymmetric condition of the Hamiltonian $H$, Eq.(\ref{Antisymmetry}) isn't hold.
Applying this cyclic evolution we can drive back the evolved quantum state to the initial state and accordingly could suppress the state from deviation.
In the light of above considerations, to control the evolution over a long time scale $t$
one can divide the total time $t$ into the number of successive cyclic evolution such that the controlled evolution $U_{c}(t)$ reads as
\be\label{Control}
U_{c}(t)=U_{cycl}(t') \overbrace{U_{cycl}(2T)U_{cycl}(2T)~.~.~.~U_{cycl}(2T)}^{\textmd{(n-1) times}}=U_{cycl}(t'),~~~t'= t- (n-1)2T,
\ee
where $(n-1)$, the number of complete cycles, is the integer part of $t/2T$ and $t=(n-1)2T+t'$.
Obviously, for $U_{c}(t)=U_{cycl}(t')$ we take the advantage of the Eq.(\ref{cyclic1}), note that the time interval of the ${n}^{th}$ cycle is: $2(n-1)T \leq t'< 2n T$.

Now we have all the ingredients to control and protect the entanglement of any given initially maximal entangled state with desired high accuracy and duration.
Our scheme to protect the entanglement in general not only can be applied for any n qubit Ising model but also for any n qubit spin system.
However in the rest of paper we restrict ourselves mostly within a three-spin Ising chain in site-dependent transverse fields to make our analysis concrete.
\subsection{Preservation of entanglement in three-qubit Ising model}
To clarify the issue as an example let us consider three-qubit Ising model in transverse field,
\be\label{Ising3}
H=J_1 \sigma_{1}^z\sigma_{2}^z + J_2 \sigma_{2}^z\sigma_{3}^z + h_2 \sigma_{2}^x,
\ee
for the sake of simplicity we set $h_1=h_3=0$.
At first we have to examine the necessary condition of Eq.(\ref{Antisymmetry}). We find that there are some unitary hermitian operators such as $A_1=\sigma_y\otimes\sigma_z\otimes\sigma_y=\sigma_1^y\sigma_2^z\sigma_3^y$, $A_2=\sigma_z\otimes\sigma_y\otimes\sigma_z=\sigma_1^z\sigma_2^y\sigma_3^z$, $A_3=I\otimes\sigma_y\otimes I=\sigma_{2}^y$, $A_4=I\otimes\sigma_y\otimes\sigma_z=\sigma_{2}^y\sigma_{3}^z$, $A_5=\sigma_x\otimes\sigma_z\otimes\sigma_y=\sigma_1^x\sigma_2^z\sigma_3^y$ and $A_6=\sigma_z\otimes\sigma_y\otimes I=\sigma_1^z\sigma_2^y$
that anti-commute with hamiltonian $H$ and satisfy the necessary condition of Eq.(\ref{Antisymmetry}). Now
we can apply the cyclic evolution, Eq(\ref{Control}), to control the dynamics and preserve the entanglement of any 3-qubit initial state.
It is worth place, to express that in \cite{Hoa} Hou and et al. present a method to preserve the entanglement of a two-qubit-spin system.\footnote[2]{They consider a two-qubit spin coupled system in the presence of Dzyaloshinskii-Moriya anisotropic anti-symmetric interaction
$$H=J_1\sigma_{1}^x\sigma_{2}^x + J_2 \sigma_{1}^y\sigma_{2}^y + D(\sigma_{1}^x\sigma_{2}^y - \sigma_{1}^y\sigma_{2}^x )$$
Although their scheme on the base of cyclic evolution is similar to Eq.(\ref{cyclic1}), it doesn't include the necessary condition of Eq.(\ref{Antisymmetry}); therefore there is no idea about the existence and determination of operator $A$. They introduce operator $O=I\otimes\sigma_z$ as a certain operator.
Note that in our notation we use operator $A$ instead of $O$. Let us reconsider this problem by our approach. The necessary condition, Eq.(\ref{Antisymmetry}), shows that there exist two unitary hermitian operators $A_1=I\otimes\sigma_z$ and $A_2=\sigma_z\otimes I$ where the hamiltonian is antisymmetric with respect to them. Therefor one can implement the cyclic evolution not only with the operator $A_1=O=I\otimes\sigma_z$ but also with the operator $A_2$ and any linear combination of $A_1$ and $A_2$.}
However our scheme not only covers and improves the results of \cite{Hoa} but also is its extension to the multi particle, n-qubit, spin systems.
Let us come back to our example and consider the dynamics of entanglement for an initially maximal entangled three-qubit GHZ state $|\psi_{_{GHZ}}\rangle=(\frac{1}{2}(|111\rangle + |000\rangle)$. To show the efficiency of the method it also would be interesting if one compare the result of entanglement dynamics for the cases of cyclic evolution and free evolution of GHZ state. Note that for free time evolution of the system during the total time $t$ the evolution operator is $U(t)=e^{-\frac{i H t}{\hbar}}$. To quantify the entanglement of system we will use the measure of concurrence vector, which is introduced in \cite{Akhtarshenas} for any m partite pure state, however for three qubit system one can use other measures such as 3-tangle or relative entropy.
In the next lines we will briefly introduce the concurrence vector following the direction of \cite{Akhtarshenas}.

\subsubsection{Concurrence vector}

\begin{figure}
 \begin{center}
  \includegraphics[width=10cm]{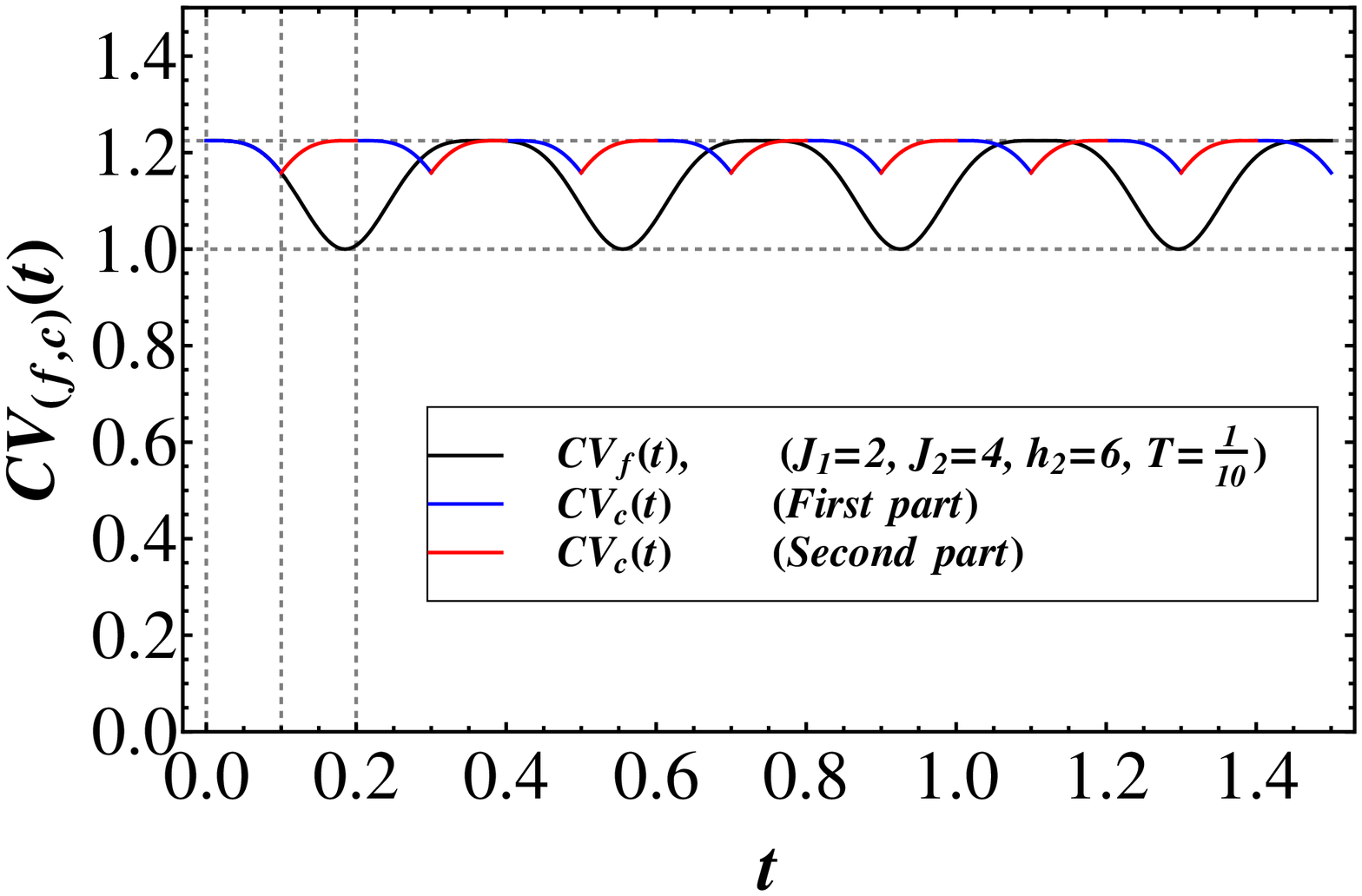}
 \caption{(Color online) black curve shows the concurrence vector of free evolution, and it is compared with controlled evolution of concurrence vector which includes sequences of falling(blue color) and rising(Red color)curves.
}
  \label{graph:C1}
 \end{center}
\end{figure}

\begin{figure}
 \begin{center}
  \includegraphics[width=10cm]{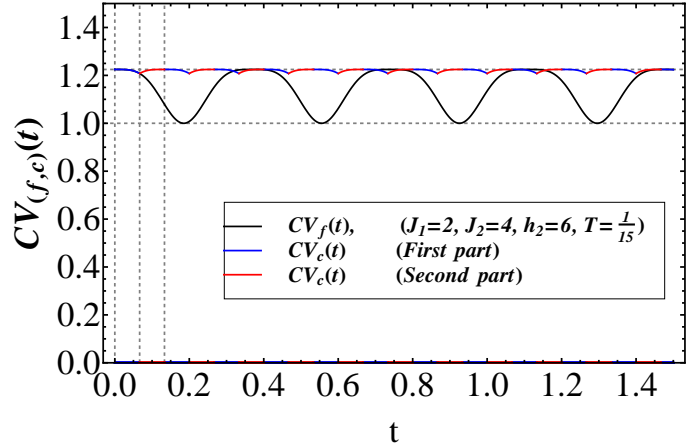}
 \caption{(Color online) black curve shows the concurrence vector of free evolution, and it is compared with controlled evolution of concurrence vector which includes sequences of falling(blue color) and rising(Red color)curves.
}
  \label{graph:C2}
 \end{center}
\end{figure}

To quantify the entanglement of two partite system, two qubit state $|\psi\rangle$, one can use the measure of concurrence \cite{Hill,Wootters}
$$
\mathcal{C}(|\psi\rangle)\equiv|\langle\psi|\sigma_{y}\otimes\sigma_{y}|\psi^{*}\rangle|
$$
where $\sigma_{y}$ is the y component of the Pauli matrices and $|\psi^{*}\rangle$ is the complex conjugate of $|\psi\rangle$.
However for a multipartite system the proper measure to quantify the entanglement is concurrence vector \cite{Akhtarshenas}.
Let us consider a m-partite pure state in the standard basis
\be
|\psi\rangle = \sum_{i_1}^{N_1}\sum_{i_2}^{N_2}...\sum_{i_n}^{N_n}{a_{{i_{_{1}}}{i_{_{2}}}...{i_{_{n}}}}}|e_{i_{_1}}\otimes e_{i_{_1}} \otimes ... \otimes e_{i_{_n}}\rangle.
\ee
Then the density matrix is $\rho=|\psi\rangle\langle\psi|$ and the norm of pairwise entanglement between the particles $i$ and $j$, in accordance with \cite{Akhtarshenas}, is defined as
\be
C^{\left\{i,j\right\}}=|\mathbf{C}^{\left\{i,j\right\}}|=\sqrt{\sum_{\alpha_{i}=1}^{N_i(N_i-1)/2}~~\sum_{\alpha_{j}=1}^{N_j(N_j-1)/2}\langle \psi|\tilde {\rho}_{{\alpha_i},{\alpha_j}}^{\left\{i,j\right\}}| \psi\rangle}
\ee
where
$\tilde {\rho}_{{\alpha_i},{\alpha_j}}^{\left\{i,j\right\}}
=M_{{\alpha_i},{\alpha_j}}^{\left\{i,j\right\}}\rho^{{T_{ij}}}M_{{\alpha_i},{\alpha_j}}^{\left\{i,j\right\}}$
and  $\rho^{{T_{ij}}}$ is the partial transpose of $\rho$ with respect to the subsystems, $i$ and $j$ and the operators $M_{{\alpha_i},{\alpha_j}}^{\left\{i,j\right\}}$ are
$$M_{{\alpha_i},{\alpha_j}}^{\left\{i,j\right\}}=I_{1} \otimes...\otimes I_{i-1}\otimes L_{\alpha_{i}}\otimes I_{i+1}\otimes...\otimes I_{j-1}\otimes L_{\alpha_{j}}\otimes I_{j+1}\otimes...\otimes I_n.$$ Note that $0\leq i < j \leq n$, $\alpha_i=1,...,N_i(N_i-1)/2$ and  $\alpha_j=1,...,N_j(N_j-1)/2$ . Also $I_k$ denotes the Identity matrix in the Hilbert space of particle $k$ and $L_{\alpha_i}$ and $L_{\alpha_j}$ represent the set of $N_i(N_i-1)/2$ and $N_j(N_j-1)/2$ generators of $SO(N_i)$ and $SO(N_j)$ groups respectively. Then according to \cite{Akhtarshenas}, the norm of concurrence vector, $\mathbf{CV}$, involves all two level entanglements shared between all particles is
\be\label{Concurrence}
CV=|\mathbf{CV}|=\sqrt{\sum_{i,j}{|\mathbf{C^{\left\{ij\right\}}}|^2}}.
\ee
To quantify the entanglement of $GHZ$ state $|\psi_{_{GHZ}}\rangle=(\frac{1}{2}(|111\rangle + |000\rangle)$,
let consider the SO(2) group where it has one generator $S= i \sigma_y$ and accordingly one can write the norm of pairwise entanglement as:
\begin{equation}\label{Concurrence componenet}
\begin{array}{cc}
C^{12}=|\mathbf{C^{12}}|=\sqrt{\langle \psi_{_{GHZ}}|M^{12}\rho^{T_{12}}M^{12}|\psi_{_{GHZ}}\rangle},~~~~M^{12} = S\otimes S\otimes I,\\
C^{13}=|\mathbf{C^{13}}|=\sqrt{\langle \psi_{_{GHZ}}|M^{13}\rho^{T_{13}}M^{13}|\psi_{_{GHZ}}\rangle},~~~~M^{13} = S\otimes I \otimes S,\\
C^{23}=|\mathbf{C^{23}}|=\sqrt{\langle \psi_{_{GHZ}}|M^{23}\rho^{T_{23}}M^{23}|\psi_{_{GHZ}}\rangle},~~~~M^{23} = I\otimes S\otimes S,
\end{array}
\end{equation}
where $\sigma_{y}$ denotes the y component of the Pauli matrices and and $I$ is a $2$ by $2$ identity matrix.
Finally, Eq.(\ref{Concurrence componenet}) together with Eq.(\ref{Concurrence}) lead to $|CV|=\sqrt{\frac{3}{2}}$ for $GHZ$ state. To study the dynamics of concurrence we consider the time evolution of $GHZ$ state. First let start with free time evolution $|\psi_{_{GHZ}}(t)\rangle=e^{-iH t}|\psi_{_{GHZ}}\rangle$, where $H$ is the Hamiltonian of Eq.(\ref{Ising3}). We find the following compact form for the concurrence vector
for the case of free evolution
\begin{equation}\label{free Concurrence}
\begin{array}{cc}
CV_f(t)= \displaystyle{\frac{\left(2 \left|2 b^4+(1-g) h_2^2 b^2+2 h_2^4\right|+\left|2 b^4+(g+7) h_2^2 b^2+2 h_2^4\right|\right)^{\frac{1}{2}}}{2 \omega^2}},\\
\end{array}
\end{equation}
the index $f$ stands for free evolution and the parameters $g$, $\omega$ and $b$ are:
\begin{equation}\label{g}
\begin{array}{cc}
g=\cos \left(4 \omega~ t\right)-4 \cos  \left(2 \omega~ t\right),\\
\omega=\sqrt{b^2+h_2^2},~~~ b=J_1+J_2.
\end{array}
\end{equation}
\\
By choosing $A=\sigma_1^y\sigma_2^z\sigma_3^y$, in the next step, we control the dynamics of the concurrence vector by implementing successive cyclic time evolution, $|\psi_{_{GHZ}}(t)\rangle=U_{c}(t)|\psi_{_{GHZ}}\rangle$, where $U_{c}(t)$ is presented by Eq.(\ref{Control}),

\begin{equation}\label{Controled Concurrence}
CV_{c}(t)=\left\{\begin{array}{cc}
\hspace{0mm}\displaystyle{\frac{\left(2 \left|2 b^4+(1-g_1) h_2^2 b^2+2 h_2^4\right|+\left|2 b^4+(g_1+7) h_2^2 b^2+2 h_2^4\right|\right)^{\frac{1}{2}}}{2 \omega^2}},\\ 2(n-1)T \leq t \leq (2n-1)T\vspace{5mm}\\
\displaystyle{\frac{\left(2 \left|2 b^4+(1-g_2) h_2^2 b^2+2 h_2^4\right|+\left|2 b^4+(g_2+7) h_2^2 b^2+2 h_2^4\right|\right)^{\frac{1}{2}}}{2 \omega^2}},\\(2n-1)T \leq t \leq 2nT\\
\end{array}\right.
\end{equation}
where the index $(n=1,2,...)$ stands for the cycle's number and the parameters $g_1(t)$, $g_2(t)$, $\omega$ and $b$ are:
\begin{equation}\label{g1g2}
\begin{array}{cc}
\hspace{0mm}g_1=\cos \left(4 \omega~ (t-2 (n-1) T)\right)-4 \cos \left(2 \omega~ (t-2 (n-1) T)\right),\\
\hspace{-17mm}g_2=\cos \left(4 \omega~ (t-2n T)\right)-4 \cos \left(2 \omega~ (t-2n T)\right),\\
\hspace{-42mm}\omega=\sqrt{b^2+h_2^2},~~~ b=J_1+J_2.\\
\end{array}
\end{equation}
We plot $CV_f(t)$ and $CV_c(t)$ in terms of $t$ with $J_1=2, J_2=4, h_2=6$ and $T=1/10$ as shown in Fig.\ref{graph:C1}. We notice that $CV_f(t)$ (black color curve) varies between ${CV_f(t)}_{max}=\sqrt{\frac{3}{2}}$ and ${CV_f(t)}_{min}={\frac{(6b^4+6h_2^4+4h_2^2b^2)^{\frac{1}{2}}}{2 \omega^2}}$ which is considerable, and $CV_f(t)_{max}$ and $CV_f(t)_{min}$ appear at the times $t_k=\frac{k\pi}{\omega}$ with $g=-3$ and $t_k=\frac{(2k+1)\pi}{2\omega}$ with $g=5$, ($k=0,1,2,...), $ respectively. One can check that the minimum value of ${CV_f(t)}_{min}$ in terms of $h_2$ appears at $h2=\pm b$ with the value $1$. In contrast to $CV_f(t)$ the situation for $CV_c(t)$ is completely different, and it almost remains near to its initial maximum value. Indeed the controlled evolution of concurrence vector includes sequences of cyclic evolutions where in each cycle we deal with two parts as in Eq.(\ref{Controled Concurrence}). Let consider $n^{th}$ cycle
with duration $2(n-1)T \leq t \leq 2 n T,~ (n=1,2,...)$. In the first part of the cycle, as it shown in Fig.\ref{graph:C1}(blue color curves), $CV_c(t)$ declines from its initial maximum value $\sqrt{\frac{3}{2}}$ to $CV_{c}(t)_{min}$, where
\be\label{Ccmin}
CV_{c}(t)_{min}=\displaystyle{\frac{({2 \left|2 b^4+(1-x) h_2^2 b^2+2 h_2^4\right|+\left|2 b^4+(x+7) h_2^2 b^2+2 h_2^4\right|})^{\frac{1}{2}}}{2 \omega^2}},
\ee

and $x=\cos (4 \omega~ T )-4 \cos (2 \omega~ T)$. It is obvious that one can keep $CV_{c}(t)_{min}$ as much as close to $\sqrt{\frac{3}{2}}$ by letting $T$ as much as close to zero. For the second part of the cycle (red color curves) $CV_c(t)$ moves up from $CV_{c}(t)_{min}$ and returns back to its initial maximum value $\sqrt{\frac{3}{2}}$. The same process goes on in the next cycles and leads to the entanglement preservation.
The comparison of $CV_f(t)$ and $CV_c(t)$ in Fig.\ref{graph:C1}(\ref{graph:C2}) for two different values of $T=1/10$ ($T=1/15$) with the same value of $J_1=2, J_2=4, h_2=6$ make it clear that to maintain the entanglement with high accuracy we have to shorten the time interval $T$ as much as possible.

\begin{figure}[H]
 \begin{center}
  \includegraphics[width=18cm]{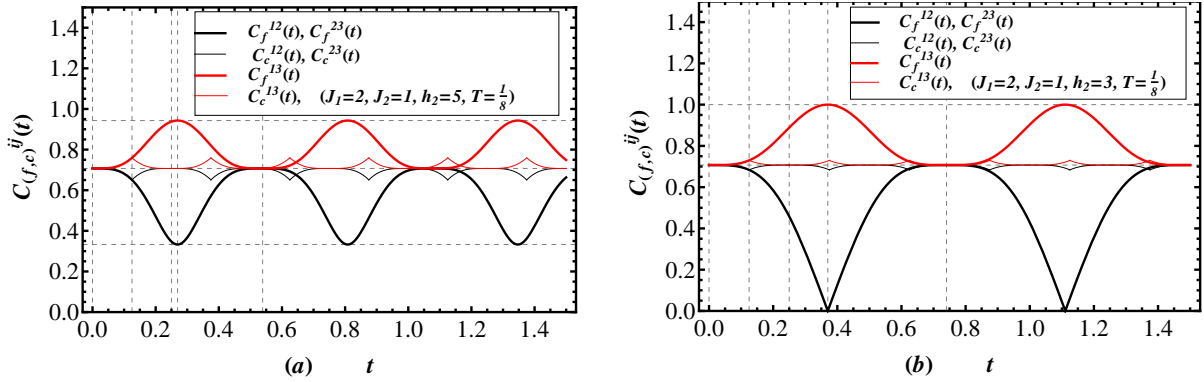}
 \caption{Fig.\ref{graph:C3}(a):(Color online) thick black (red) curves show the pairwise entanglements $C^{12}_f(t)=C^{23}_f(t)$ ($C^{13}_f(t)$) and thin black (red) curves show the pairwise entanglements $C^{12}_c(t)=C^{23}_c(t)$ ($C^{13}_f(t)$) for general case $(J_1=2, J_2=1, h_2=5)$ with $T=1/8$ respectively. As it is seen from Fig.\ref{graph:C3}(a,b)while $C^{ij}_f(t)s$ vary between their upper and lower bounds, $C^{ij}_c(t)s$ almost remain near to the value $(\frac{\sqrt{2}}{2})$ which is the upper (lower) bound of $C^{12}_f(t)=C^{23}_f(t)$ ($C^{13}_f(t)$). The first, second and third vertical grid lines represent the times $(t=0, t=1/8,  t=2/8 )$ however the forth and fifth ones represent the times $(t=\pi/2\omega$ and  $t=\pi/\omega )$ respectively. The horizontal grid lines shows the upper and lower bounds for $C^{12}_f(t)=C^{23}_f(t)$ and $C^{13}_f(t)$. Fig.\ref{graph:C3}(b):(Color online) thick black (red) curves show the pairwise entanglements $C^{12}_f(t)=C^{23}_f(t)$ ($C^{13}_f(t)$) and  thin black (red) curves show the pairwise entanglements $C^{12}_c(t)=C^{23}_c(t)$ ($C^{13}_f(t)$) for special case $(J_1=2, J_2=1, h_2=3)$ with $T=1/8$ respectively. The vertical grid lines have the same explanation as in Fig.\ref{graph:C3}(a). The horizontal grid lines shows the upper and lower bounds of $C^{12}_f(t)=C^{23}_f(t)$ and $C^{13}_f(t)$. At the times $(t=(2k+1)\pi/2\omega, ~k=0,1,2,...)$  $C^{12}_f(t)=C^{23}_f(t)$ ($C^{13}_f(t)$) achieve $0$$(1)$ respectively. That is at this times the particles $(1,2)$ and $(2,3)$ are separable while particles $(1,3)$ are maximally entangled. Also comparing Fig.\ref{graph:C3}(a,b) makes it clear that the upper(lower) bound of $C^{12}_f(t)=C^{23}_f(t)$ ($C^{13}_f(t)$) are the same for the special and the general cases and is a constant value $\sqrt{2}/2$.
  }
  \label{graph:C3}
 \end{center}
\end{figure}

So far we have just considered the time evolution of concurrence vector (multiparty entanglement) for three-qubit GHZ state, one may interested in to control the pairwise entanglement between particles $i$ and $j$ in three-qubit GHZ state. To this end, we recall Eq.(\ref{Concurrence componenet}) and for free evolution we find the following results for the pairwise entanglement:
\begin{equation}\label{free pairwiseConcurrence}
C^{12}_f(t)=C^{23}_f(t)= \displaystyle{\frac{\sqrt{2 b^4+(1-g) h_2^2 b^2+2 h_2^4}}{2 \omega^2}},~~
C^{13}_f(t)= \displaystyle{\frac{\sqrt{2 b^4+(g+7) h_2^2 b^2+2 h_2^4}}{2 \omega^2}},
\end{equation}
and for controlled evolution we get
\begin{equation}\label{Controled pairwiseConcurrence}
\left\{\begin{array}{cc}
\hspace{0mm}C^{12}_c(t)=C^{23}_c(t)= \displaystyle{\frac{\sqrt{2 b^4+(1-g_1) h_2^2 b^2+2 h_2^4}}{2 \omega^2}},~~
C^{13}_c(t)= \displaystyle{\frac{\sqrt{2 b^4+(g_1+7) h_2^2 b^2+2 h_2^4}}{2 \omega^2}},\\ 2(n-1)T \leq t \leq (2n-1)T\vspace{5mm}\\
C^{12}_c(t)=C^{23}_c(t)= \displaystyle{\frac{\sqrt{2 b^4+(1-g_2) h_2^2 b^2+2 h_2^4}}{2 \omega^2}},~~
C^{13}_c(t)= \displaystyle{\frac{\sqrt{2 b^4+(g_2+7) h_2^2 b^2+2 h_2^4}}{2 \omega^2}},\\(2n-1)T \leq t \leq 2nT\\
\end{array}\right.
\end{equation}
where $g$, $g_1$ and $g_2$ are given as in Eq.(\ref{g}) and Eq.(\ref{g1g2}). Obviously one can check that at the times $(t_k=\frac{k\pi}{\omega}, k=0,1,2,...)$ where $g=-3$ for any values of $h_2$ and $b$, $C^{12}_f(t)=C^{23}_f(t)=C^{13}_f(t)=\frac{\sqrt{2}}{2}$ which verify that $CV_f(t)_{max}=\frac{\sqrt{3}}{2}$. On the other hand at the times $(t_k=\frac{(2k+1)\pi}{2\omega}, k=0,1,2,...)$ where $g=5$ we find that $C^{12}_f(t)=C^{23}_f(t)=\displaystyle{\frac{\mid h_2^2-b^2\mid}{\sqrt{2}\omega^2}}$ and $C^{13}_f(t)=\displaystyle{\frac{\sqrt{2b^4+12h_2^2b^2+2h_2^2}}{2\omega^2}}$, where they verify the value of ${CV_f(t)}_{min}=\displaystyle{{\frac{(6b^4+6h_2^4+4h_2^2b^2)^{\frac{1}{2}}}{2 \omega^2}}}$. Therefore, the pairwise entanglement $C^{12}_f(t)=C^{23}_f(t)$ and  $C^{13}_f(t)$ vary periodically  between $ \displaystyle{\frac{\mid h_2^2-b^2\mid}{\sqrt{2}\omega^2}} \leq (C^{12}_f(t)=C^{23}_f(t))\leq \displaystyle{\frac{\sqrt{2}}{2}}$ and $ \displaystyle{\frac{\sqrt{2}}{2}} \leq C^{13}_f(t)\leq \displaystyle{\frac{\sqrt{2b^4+12h_2^2b^2+2h_2^2}}{2\omega^2}}$ respectively. In fact the upper bound of $C^{12}_f(t)=C^{23}_f(t)$ and the lower bound of $C^{13}_f(t)$ are the same and equal to $(\frac{\sqrt{2}}{2})$. On the other hand, the lower bound of $C^{12}_f(t)=C^{23}_f(t)$ and the upper bound of $C^{13}_f(t)$  generally are functions of $(h_2, b)$, however for the special case $h_2=\pm b$ we get $C^{12}_f(t)=C^{23}_f(t)=0$ and $C^{13}_f(t)=1$. In Fig.\ref{graph:C3}  we plot $C^{12}_{(f,c)}(t)=C^{23}_{(f,c)}(t)$ and $C^{13}_{(f,c)}(t)$ for the special case $h_2= b$  and the general case $h_2\neq b$  with $(J_1=2, J_2=1, h_2=5,T=\frac{1}{8})$ to compare the results.

\begin{figure}[H]
 \begin{center}
  \includegraphics[width=10cm]{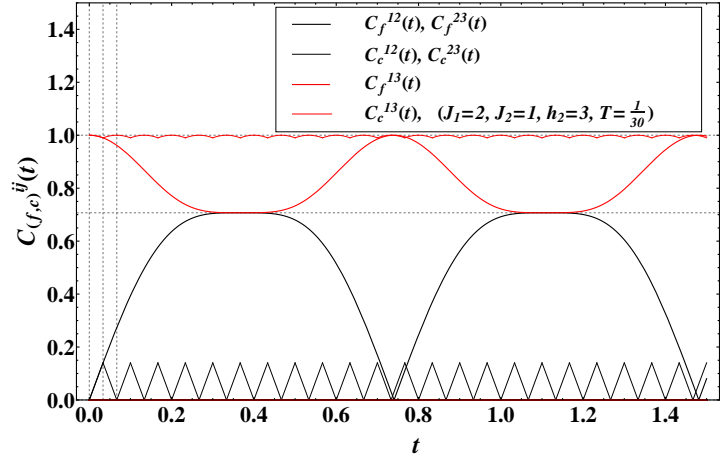}
 \caption{(Color online) Comparison of pairwise concurrences with parameters $J_1=2, J_2=1, h_2=3, T=\displaystyle{\frac{1}{30}}$. Thick black(red) curves show the pairwise concurrence of free evolution $C^{12}_f(t)=C^{23}_f(t)$ $(C^{13}_f(t))$ and thin black(red) curves show the pairwise concurrence of controlled evolution $C^{12}_c(t)=C^{23}_c(t)$ $(C^{13}_c(t))$. Therefor particles $(1,2)$ and $(2,3)$ are separable while particles $(1,3)$ are maximally entangled.
  }
  \label{graph:C4}
 \end{center}
\end{figure}

Now to preserve the pairwise entanglement $C^{ij}(t)$ on a desired value, first we let the state evolve freely till to reach that value, and then we use  controlled time evolution of pairwise entanglement ($C^{ij}_c(t)$). As an example let us consider the case ($C^{12}_f(t)=C^{23}_f(t)=0$, $C^{13}_f(t)=1$),  i.e., the particles $(1,2)$ and $(2,3)$ are separable while particles $(1,3)$ are maximally entangled. In order to keep the pairwise entanglements on these values, as it clear from Fig.3(b), we just need to shift the time $t$ by $\displaystyle{\frac{\pi}{2\omega}}$ in Eq.(\ref{Controled pairwiseConcurrence}). Fig.\ref{graph:C4} shows the comparison of the free and controlled evolution of pairwise concurrences. The parameters are  $J_1=2, J_2=1, h_2=3, T=\displaystyle{\frac{1}{30}}$.

\section{Conclusion}
In conclusion we have proposed a method based on the antisymmetric property of the hamiltonian to control the dynamics of the entanglement and hence to protect the multiparty and pairwise entanglement of three-qubit GHZ state as an initial state which governed by 3-particle Ising chain.  We show that the control process consists of a sequences of cyclic evolution where each cycle took place at the time $2T$. We find that to preserve the entanglement with any high degree of accuracy one has to shorten the time interval $T$ as much as possible.

\end{document}